\documentclass[namedreferences]{solarphysics}
\usepackage[optionalrh,solaromanenum]{spr-sola-addons} 
\usepackage{graphicx}                    
\usepackage{color}                       
\usepackage{url}                         

\newcommand{\etal}{{\it et al.}}

\newcommand{\adv}{    {\it Adv. Space Res.}}
\newcommand{\annG}{   {\it Ann. Geophys.}}
\newcommand{\aap}{    {\it Astron. Astrophys.}}

\newcommand{\apj}{    {\it Astrophys. J.}}
\newcommand{\apjs}{   {\it Astrophys. J. Suppl.}}
\newcommand{\apjl}{   {\it Astrophys. J. Lett.}}
\newcommand{\apss}{   {\it Astrophys. Space Sci.}}

\newcommand{\grl}{    {\it Geophys. Res. Lett.}}

\newcommand{\jgr}{    {\it J. Geophys. Res.}}
\newcommand{\mnras}{  {\it Mon. Not. Roy. Astron. Soc.}}

\newcommand{\solphys}{{\it Solar Phys.}}

\newcommand{\ssr}{    {\it Space Sci. Rev.}}

\begin{document}

\begin{article}

\begin{opening}

\title{Validation of Spherically Symmetric Inversion by Use of a Tomographic 
Reconstructed Three-Dimensional Electron Density of the Solar Corona} 
%
\author{Tongjiang~\surname{Wang}$^{1,2}$\sep
         Joseph M.~\surname{Davila}$^{2}$ \\
(submitted 2013 October 24;  accepted 2014 May 12)     
       }
%
 \runningauthor{Wang \& Davila}
 \runningtitle{Validation of Spherically Symmetric Inversion}

%
  \institute{$^{1}$ Department of Physics, Catholic University of America,
   620 Michigan Avenue NE, Washington, DC 20064, USA 
                     email: \url{tongjiang.wang@nasa.gov}\\
              $^{2}$ NASA Goddard Space Flight Center, Code 671, Greenbelt, MD 20770, USA
             }

\begin{abstract}
Determination of the coronal electron density by the inversion of white-light polarized 
brightness ($pB$) measurements by coronagraphs is a classic problem in solar physics. 
An inversion technique based on the spherically symmetric geometry (Spherically Symmetric Inversion,
SSI) was developed in the 1950s, and has been widely applied to interpret various observations. 
However, to date there is no study about uncertainty estimation of this method.
In this study we present the detailed assessment of this method using a three-dimensional (3D) 
electron density in the corona from 1.5 to 4 $R_\odot$ as a model, which is reconstructed by 
tomography method from STEREO/COR1 observations during solar minimum in February 2008
(Carrington rotation, CR 2066). We first show in theory and observation that 
the spherically symmetric polynomial approximation (SSPA) method and the Van de Hulst 
inversion technique are equivalent. Then we assess the SSPA method using synthesized 
$pB$ images from the 3D density model, and find that the SSPA density values are close to 
the model inputs for the streamer core near the plane of the sky (POS) with 
differences generally less than a factor of two or so; the former has the lower peak 
but more spread in both longitudinal and latitudinal directions than the latter. 
We estimate that the SSPA method may resolve the coronal density structure near the POS 
with angular resolution in longitude of about 50$^{\circ}$. Our results confirm 
the suggestion that the SSI method is applicable to the solar minimum streamer (belt) 
as stated in some previous studies. In addition, we demonstrate that the SSPA method 
can be used to reconstruct the 3D coronal density, roughly in agreement with that 
by tomography for a period of low solar activity (CR 2066). We suggest that the SSI method 
is complementary to the 3D tomographic technique in some cases, given that 
the development of the latter is still an ongoing research effort.
\end{abstract}

%

\end{opening}

%
\section{Introduction}
The electron density is a fundamental parameter in plasma physics. Knowledge of the
three-dimensional (3D) electron density structure is very important for our understanding 
of physical processes in the solar corona, such as the coronal heating and the 
acceleration of the solar wind ({\it e.g.}, \opencite{mun77}; \opencite{cra99}). 
The density structure of the corona strongly affects the propagation of CMEs 
\cite{ods99, ril01, ods02, man04}.
The density is also important for estimates of the Alfv\'{e}n Mach number and compression
rate of CME-driven shocks \cite{rea99, sok04, man05}, and for the interpretation 
of solar radio emission such as type II and type IV radio bursts produced by coronal eruptions 
\cite{car04, cho07, she13, ram13, zuc14}. 

The K corona arises from Thomson scattering of photospheric white light from free electrons 
({\it e.g.}, \opencite{bil66}). Because the emission is optically thin, the measured 
signal is a contribution from electrons all along the line of sight (LOS). The derivation of 
the electron density in the K corona from the total brightness ($B$) or polarized radiance 
($pB$) is a classical problem of coronal physics, first addressed by \inlinecite{min30} and
 \inlinecite{van50}. Because of difficulties in generally separating the K-coronal component 
from the F-coronal component arising from interplanetary dust scattering ({\it e.g.}, 
\opencite{bil66}; \opencite{hay01}), most of the inversion techniques are practical for 
$pB$ measurements. The F-coronal polarization is not very well understood (see reviews
by \opencite{kou85}; \opencite{kim98}). It is generally accepted that 
the polarized contribution of the F corona can be ignored 
within 5 $R_{\odot}$ \cite{man92, kou85, hay01}, but some observations show that
the F corona is almost unpolarized even at elongations ranging from 10 to 16 $R_{\odot}$
(\opencite{bla66a}, \citeyear{bla66b}; \opencite{bla67}). In addition, the F corona 
dominates the total brightness of
the corona beyond about 4 or 5 $R_{\odot}$ and make it difficult to recover the much
fainter K-corona emission \cite{sai77, kou85, hay01}. To retrieve the
electron density of the corona from a single 2D $pB$-image, one needs to assume
some special geometries for the distribution of electrons along the LOS. The previous studies
have modeled the electron density distribution in several ways, including the simple
spherically symmetric model \cite{van50}, the axisymmetric model \cite{sai70, mun77, que02}, 
or the models that take into account large-scale structures, such as polar plumes in 
the coronal holes \cite{bar08}, or active streamers in the equatorial regions \cite{guh96}. 

\begin{figure} 
\centerline{\includegraphics[width=0.8\textwidth,clip=]{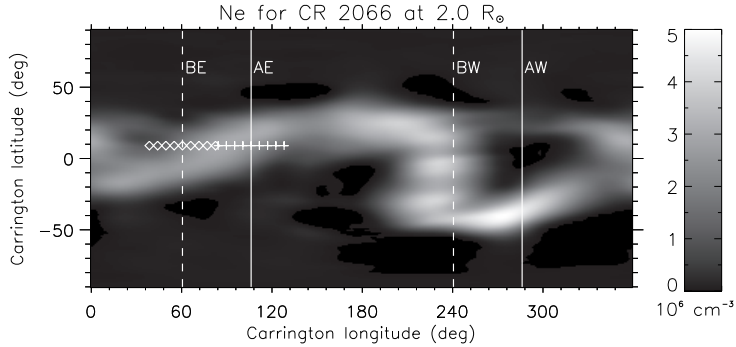}}
\caption{Spherical cross section of the tomographic reconstructed 
3D coronal electron density for CR 2066 at a heliocentric distance of 2.0 $R_{\odot}$.
Two vertical solid lines marked with AE and AW indicate the positions of 
the east and west limbs in the COR1-A image observed at 12:00 UT on 
8 February 2008, and those dashed lines with BE and BW the positions of 
east and west limbs in the COR1-B image at the same time. The symbols
marked {\it diamonds} and {\it pluses} indicate the locations for the SSPA inversions
shown in Figures~\ref{fig:neva} and~\ref{fig:nevb}.  } 
\label{fig:necr}
\end{figure}

Among the above inversion methods, the spherically symmetric inversion (SSI) developed
by \inlinecite{van50} (called the {\it Van de Hulst} inversion thereafter), was the most 
representative and commonly used. He found that the density integral for $pB$ signals 
becomes invertible if the latitudinal and azimuthal gradients in electron density are 
weaker than the radial gradient ({\it i.e.}, a local spherical symmetry approximation). 
This classic inversion technique has been applied to establish the standard density models 
of the coronal background at equator and pole in the solar minimum and maximum 
(see \opencite{all00}) and the density models of near-symmetric coronal structures 
such as streamers and coronal holes \cite{sai77, gib95, guh95, guh96, gib99}. 
The SSI method was also used to analyze detailed density distribution of fine coronal 
structures observed in eclipses ({\it e.g.}, \opencite{kou94}; \opencite{nov96}), and
to derive the 2D density distribution of the entire corona \cite{hay01, que02}, when
the spherical symmetry is assumed holding locally. The importance of the SSI method for
coronal density determination has been demonstrated by wide applications of the derived 
densities such as in testing
models of the acceleration mechanism of the fast solar wind \cite{que07, lal10},
interpreting sources of type II and type IV radio bursts \cite{cho07, she13, ram13, zuc14},
and determining the coronal magnetic field strengths from fast magnetosonic waves by 
global coronal seismology \cite{kwo13}.

However, in contrast to extensive applications, studies on evaluation of the SSI method 
are few in the literature. \inlinecite{gib99} compared the white light 
densities to those determined from the density-sensitive EUV line ratios of 
Si\,{\sc{ix}} 350/342 \AA\
observed by SOHO/CDS, and found that densities determined from these two different 
analysis techniques match extremely well in the low corona for a very symmetric solar
minimum streamer structure. Similarly, \inlinecite{lee08} compared densities of various
coronal structures determined by inverting MLSO MK4 $pB$ maps and from the line ratios of
O\,{\sc{vi}} 1032/1037.6 \AA\ observed by SOHO/UVCS, and found that the mean densities in
a streamer by the two methods are consistent, while the coronal densities for a coronal
hole and an active region are within a factor of two. These results are
encouraging, and suggest that the 2D white-light density distribution in coronal 
structures can be very useful for other studies, but a detailed assessment is required
for its better application, such as information about the limitation of the SSI method and 
the uncertainty of derived densities.  To achieve this goal, one may use 
synthetic $pB$ images from 3D densities of the corona reconstructed by tomographic techniques 
(\opencite{fra02}; \opencite{fra07}, \citeyear{fra10}; \opencite{vas08}; \opencite{kra09}; 
\opencite{but10}; \opencite{bar13}), or simulated by global 3D MHD models 
\cite{mik99, lin99, usm00, gro00, hay05, ril06, fen07, hu08, lio09, van14}. 
 
\begin{figure}
\centerline{\includegraphics[width=1.\textwidth,clip=]{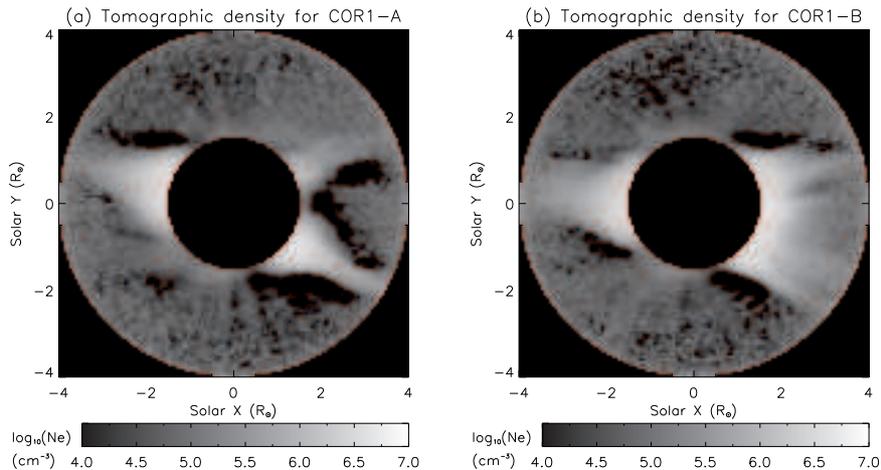}}
\caption{Cross sections of the tomographic 3D coronal density in the plane of the sky (POS), 
corresponding to positions of COR1-A (a) and COR1-B (b) at 12:00 UT on 8 February 2008. }
\label{fig:nemap}
\end{figure}

The tomographic technique is a sophisticated method which reconstructs optically thin 3D coronal 
density structures using observations from multiple viewing directions. The use of this method 
in solar physics was previously proposed by \inlinecite{dav94}, and later this method has been 
applied to the SOHO/LASCO \cite{fra02} and STEREO/COR1 data \cite{kra09, kra14}.
For a solar coronal tomography based on observations made by a single spacecraft
or only from the Earth-based coronagraph, data typically need to be gathered over a period
of half solar rotation, so, generally, only structures that are stationary over about two weeks 
can be reliably reconstructed. Therefore, this technique is not applicable to eclipses or, 
perhaps, periods of high level of solar activity, although the 3D coronal electron density 
can be routinely computed \cite{but05, kra09}. Similarly, as global MHD models of the
corona need measurements of photospheric magnetic field data over a solar rotation, it is also 
difficult using the MHD method to reconstruct dynamic or rapidly-evolving coronal structures 
matching to observations. Thus, the SSI analysis could be very useful in some cases when 
the tomography is not suitable. In addition, the SSI method is also useful in order to 
investigate the coronal density variability over a long term period (several solar cycles) 
when modern quality synoptic observations were not available and relate it to 
the modern state of the art reconstructions.

In this study, we choose the 3D coronal density obtained by tomography as a model
in order to estimate uncertainties of the SSI method. \inlinecite{vas08} compared 
the tomographic reconstruction and a 3D MHD model of the corona, and found that at lower 
heights the MHD models have better agreement with the tomographic densities in the region 
below 3.5 $R_{\odot}$, but become more problematic at larger heights. 
They also showed that the tomographic reconstruction has 
more smaller-scale structures within the streamer belt than the model can reproduce. 
Moreover, the tomographic reconstruction is entirely based on {\it coronal} observations, 
while the MHD models are primary based on photospheric boundary conditions.
This suggests that the tomographic reconstructions are more realistic and thus may be 
more suitable to be used as a model in order to evaluate uncertainties of the SSI method. 
  
This article is organized as follows. Section~\ref{sctssi} describes two SSI methods 
and their relationship. Section~\ref{sctrlt} presents the evaluation of the SSI 
method. We demonstrate the 3D density reconstruction by the SSI method based on
real data in Section~\ref{sctd3d}. The discussion and conclusions as well as 
the potential extension of our work are given in Section~\ref{sctdc}.

\begin{figure}
\centerline{\includegraphics[width=1.\textwidth,clip=]{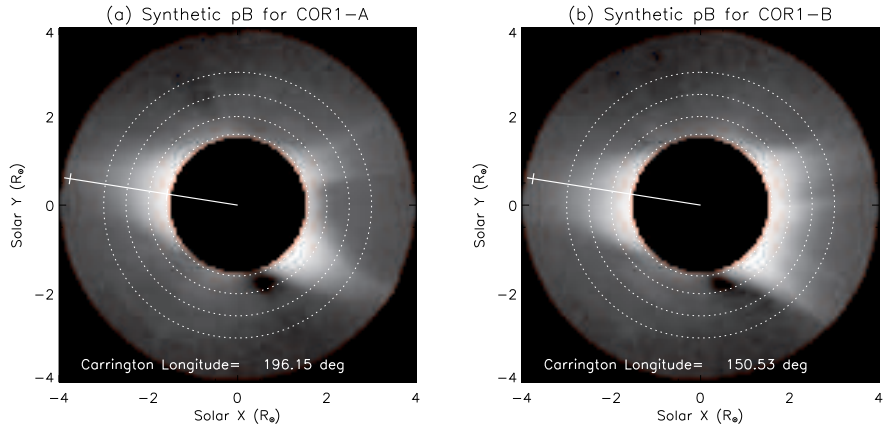}}
\caption{Synthetic $pB$ images based on the 3D coronal density model reconstructed 
by tomography. (a) Viewed from COR1-A, and (b) from COR1-B 
at 12:00 UT on 8 February 2008. Carrington longitudes of the viewing 
direction are marked at the bottom of the images. Solid lines mark the
positions where the $pB$ intensity profiles shown in Figure~\ref{fig:pblc} are 
extracted. Short bars with a size of 0.31 $R_{\odot}$ show the scale,
over which the $pB$ intensity is averaged across the radial cut. Four circles 
(dotted lines) at 1.6, 2.0, 2.5 and 3.0 $R_{\odot}$ mark the paths along which
the density profiles are shown in Figure~\ref{fig:neap}.}
\label{fig:pbmap}
\end{figure}

\begin{figure}
\centerline{\includegraphics[width=0.8\textwidth,clip=]{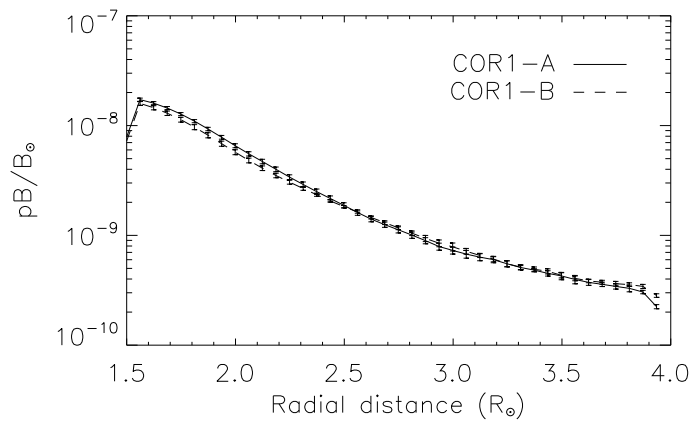}}
\caption{Radial profiles of $pB$  for COR1-A (solid line)
and COR1-B (dashed line), extracted along the radial cuts shown in
Figure~\ref{fig:pbmap}. The error bars are the standard deviations for average
over five pixels across the cut. }
\label{fig:pblc}
\end{figure}

\section{Spherically Symmetric Inversion Method}
\label{sctssi}

Following derivations in \inlinecite{bil66}, $pB$ at a point, P, 
on the plane of sky (POS) can be expressed as,
\begin{equation}
pB (\rho) =\frac{B_{\odot}}{1-u/3}\left(\frac{3\sigma_T}{16}\right)
          \int_{LOS}[(1-u)A(r)+uB(r)]\frac{\rho^2}{r^2} N(s)ds, \label{eqpb}
\end{equation}
where the POS is defined as a plane crossing the Sun's center and perpendicular to the LOS,
$N$ is the electron density, $\rho$ is the perpendicular distance between 
the LOS and Sun center, $r$ is the radial distance from the Sun center, 
and $s$ is the distance measured from P along the LOS. These distances
are related by $r^2=\rho^2+s^2$. $B_{\odot}$ is the mean solar brightness. 
$\sigma_T=\frac{8}{3}\pi{r}_0^2=6.65\times10^{-25}$ cm$^2$, is the Thompson scattering cross section 
for a single electron, where ${r}_0$ is the classical electron radius. Note that 
$r_0^2=7.94\times10^{-26}$ cm$^2$ is the ``Thompson scattering cross section" as referred to
in \citeauthor{bil66} \shortcite{bil66}. $u$ is the limb darkening coefficient. 
In addition, $A(r)$ and $B(r)$ are geometrical factors given by 
\citeauthor{bil66} \shortcite{bil66}
\begin{eqnarray}
 A(r) &  = & {\rm cos}\Omega\,{\rm sin}^2\Omega, \label{eqar} \\
 B(r) &  = & -\frac{1}{8}\left[1-3\,{\rm sin}^2\Omega-{\rm cos}^2\Omega\left(\frac{1+3\,{\rm sin}^2\Omega}{{\rm sin}\Omega}\right){\rm ln}\left(\frac{1+{\rm sin}\Omega}{{\rm cos}\Omega}\right)\right], \label{eqbr}
\end{eqnarray}
where the angle $\Omega$ is defined by ${\rm sin}\Omega=R_{\odot}/r$.

If electron density is a function of $r$ only, Equation (\ref{eqpb}) can be written in the form
\begin{equation} 
  pB(\rho) =C \int_{\rho}^{+\infty}[(1-u)A(r)+uB(r)]N(r)\frac{\rho^2 dr}{r\sqrt{r^2-\rho^2}}, 
     \label{eqpbr}
\end{equation}
where $r$ and $\rho$ are in units of $R_{\odot}$, and $C=(3/8)\sigma_TR_{\odot}B_{\odot}/(1-u/3)$. \inlinecite{van50} developed a technique for the inversion of $pB$ measurements (called the {\it Van de Hulst} inversion). To implement this technique, one needs first to fit the $pB$ data using a curve
in the polynomial form, specifically, 
\begin{equation}
 pB(\rho) = \sum_k a_k \rho^{-k}. \label{eqpbsum}
\end{equation}
Once the coefficients $a_k$ are determined, the solution of Equation (\ref{eqpbr}) is then given by
\begin{equation}
 N(r) = \frac{\sum_k (a_k/c_{k+1})r^{-(k+1)}}{C((1-u)A(r)+uB(r))}, \label{eqnr}
\end{equation}
where 
\begin{equation}
c_k=\int_0^{\pi/2}{\rm cos}^k{\phi}\,d\phi = 
   \frac{\sqrt{\pi}}{2}\frac{\Gamma((k+1)/2)}{\Gamma(k/2+1)},
\end{equation}
where $\Gamma$ is the gamma function. Note that the constant $C$ here is different from
that in \inlinecite{van50}, because we have used the expression of $pB$ in \inlinecite{bil66}. 
More generally, the index $k$ of radial power law in Equation~(\ref{eqpbsum}) is real 
but not necessarily integer valued. This leads to the modified Van de Hulst technique assuming
$pB(\rho)=\sum_k{a_k}\rho^{-d_k}$, where $a_k$ and $d_k$ are the fit coefficients, then
$c_{k+1}$ and $r^{-(k+1)}$ in Equation (\ref{eqnr}) need to be replaced with
$c_{d_k+1}$ and $r^{-(d_k+1)}$ ({\it e.g.}, \opencite{sai77}; \opencite{guh96}; \opencite{gib99}).

\inlinecite{hay01} developed another SSI technique by assuming the radial electron
density distribution in the polynomial form, $N(r)=\sum_k b_k r^{-k}$, and have 
used it to the inversion of total brightness observations. We here adopt this technique for
the $pB$ inversion, and call it the spherically symmetric polynomial approximation (SSPA) method
for short. By substituting the polynominal for $N(r)$ in Equation (\ref{eqpbr}), we obtain 
\begin{equation}
 pB (\rho) = \sum_k b_k G_k(\rho), \label{eqpbg}
\end{equation}
where
\begin{equation}
 G_k(\rho) = C\int_{\rho}^{+\infty}[(1-u)A(r)+uB(r)](r^{-k})\frac{\rho^2 dr}{r\sqrt{r^2-\rho^2}}, \label{eqgkr}
\end{equation}
For easier calculation, using the substitution $r=\rho/{\rm cos}\theta$ the integral can 
be transformed into
\begin{equation} 
 G_k(\rho) = C\rho^{-k+1}\int_0^{\pi/2}[(1-u)A(r(\theta))+uB(r(\theta))]{\rm cos}^k\theta\,d\theta, \label{eqgktht}
\end{equation}
where $\theta$ is the angle from the POS to a direction from the Sun center to a point of 
distance $s$ along the LOS. Since the integral $G_k(\rho)$ can be numerically calculated for 
all desired impact distances $\rho$ and exponents $k$, substituting the observed $pB$ 
curve along a radial trace for the left-hand side of Equation (\ref{eqpbg}) 
the only unknowns are the coefficients $b_k$. We determine the coefficients by a multivariate 
least-squares fit to the curve of $pB(\rho)$ (using the function of 
{\it svdfit} provided by IDL, Interactive Data Language). The radial distribution 
of electron density is then obtained by substituting the resulting coefficients directly into 
the polynomial form. To select appropriate degrees of polynominal we did some experiments 
using coronal density models 
for the region between 1.5 and 6 $R_{\odot}$ in \inlinecite{sai77}, and found that choosing 
the first five terms ($k=1-5$) can determine $N(r)$ with the relative errors within 5\% 
and reproduce $pB$ measurements with the relative errors within 1\%. We thus use the 5-degree 
polynomial fits for all the SSPA inversions\footnote{IDL codes of the SSPA method for inversions of
COR1 pB data in 1D and 2D are available for download (http://solar.physics.montana.edu/wangtj/sspa.tar)} 
in our study. 

To look into the relationship between the SSPA and Van de Hulst inversions
theoretically, we use Taylor series to approximate the functions 
$A(r)$ and $B(r)$ in the case when $r>1$ with $r$ in unit of $R_\odot$,
\begin{eqnarray} 
A(r) &\approx&  \left(\frac{1}{r}\right)^2-\frac{1}{2}\left(\frac{1}{r}\right)^4-\frac{1}{8}\left(\frac{1}{r}\right)^6+ ...\, , \\
B(r) &\approx&  \frac{2}{3}\left(\frac{1}{r}\right)^2-\frac{4}{15}\left(\frac{1}{r}\right)^4-\frac{3}{40}\left(\frac{1}{r}\right)^6+...\,.  \\
\end{eqnarray}
For the Taylor polynomial approximations of degree six above, the relative errors are 
less than 1\% when $r>1.5$. Since $A(r)$ and $B(r)$ are on the order of $O(1/r^2)$ for 
very large $r$, by keeping only the terms of $1/r^2$ we have $(1-u)A(r)+uB(r)\approx(1-u/3)(1/r^2)$,
 which corresponds to the point source approximation \cite{fra10}. Applying this
approximation to Equation~(\ref{eqnr}) for the Van de Hulst inversion, we obtain
\begin{equation}
  N(r) = \frac{1}{C(1-u/3)}\sum_k\left(\frac{a_k}{c_{k+1}}\right)r^{-(k-1)}. \label{eqnrapx}
\end{equation}
This verifies the polynomial form of $N(r)$ assumed in the SSPA method. Conversely, 
if $N(r)$ is given in the same form as Equation~(\ref{eqnrapx}), likewise we can recover 
$pB(\rho)=\sum_k a_k \rho^{-k}$ as assumed in the Van de Hulst inversion from the SSPA inversion 
using Equations~(\ref{eqpbg}) and~(\ref{eqgktht}) with the point source approximation 
for $(1-u)A(r)+uB(r)$. Thus, we have proved that the SSPA and Van de Hulst inversions 
are identical on the order of $O(1/r^2)$, and their difference in higher orders can be reduced 
by increasing the degrees of $k$. Note that the above analysis is also valid for the case 
when $k$ is real. In Section~\ref{sctstrm}, we will show that the coronal densities determined
from COR1 images by the SSPA and Van de Hulst inversions with $k$=5 are almost identical. 
Therefore, the assessment of the SSPA inversion in the following is equivalent to that of 
the Van de Hulst inversion.
                
\begin{figure}
\centerline{\includegraphics[width=1.\textwidth,clip=]{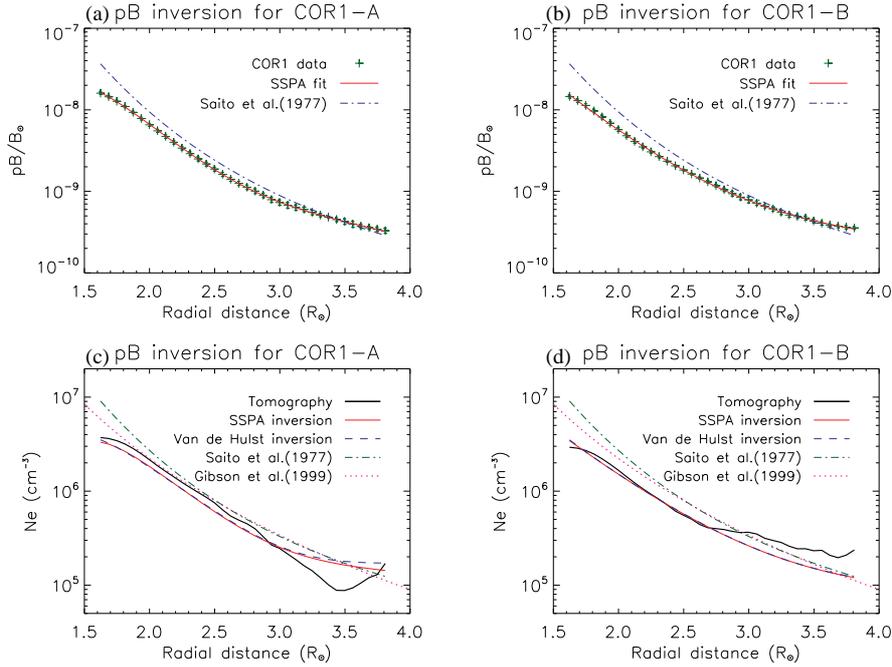}}
\caption{(a) and (b): The synthetic $pB$ data (denoted with {\it pluses}) along a radial cut 
shown in COR1-A and -B images in Figure~\ref{fig:pbmap}, and their least-squares fit (solid line) 
using the SSPA method. (c) and (d): The density profiles derived with the SSPA inversion
(thin solid line) and the Van de Hulst inversion (dashed line), as well as the model 
densities in the POS (thick solid line). Note that in panel (d) the density profiles obtained by
the SSPA and the Van de Hulst inversions are overlaid completely. For comparison, 
the $pB$ profile and the density model for the equatorial background derived by 
Saito, Poland, and Munro (1977) from the {\it Skylab} data are plotted with dot-dashed lines. 
In (c) and (d) the overplotted curves also include the solar minimum streamer densities 
(dotted line) obtained by Gibson \etal~(1999) from Mauna Loa/LASCO C2 data. }
\label{fig:pbinv}
\end{figure}

\section{Evaluation of the SSI Method}
\label{sctrlt}

\subsection{Coronal Density Model for Evaluation of SSI Method} 
\label{scttt}
As a model for evaluation of SSI method we used a 3D coronal electron density obtained by
tomography method applied to white-light coronagraph data \cite{kra09}. The data were
acquired by the inner coronagraph (COR1) telescopes aboard the twin {\it Solar Terrestrial Relation
Observatory} (STEREO) spacecraft. COR1 observes the white-light K corona from about 1.4 to 4
$R_\odot$ in a waveband of 22.5 nm wide centered on H$\alpha$ line at 656 nm with a time cadence of 5
minutes. Regularized tomographic inversion method by \inlinecite{kra09} with the limb darkening
coefficient $u$=0.6 provided the reconstruction of a 3D coronal electron density for the period 
of 2008 February 1--14 (consisting of 28 $pB$ images from COR1-B) that corresponds to Carrington
Rotation (CR) 2066. The scattered light in the $pB$ data was removed by subtracting a combination of
the monthly minimum and the roll minimum backgrounds \cite{tho10}. The reconstruction
domain is a $128^3$ rectangular grid covering a spherical region between 1.5 and 4 $R_\odot$. We
evaluate the SSPA inversion method using this 3D density reconstruction as a model.

In order to synthesize $pB$ images observed by COR1-A and -B at a given time,
we first determine the orbital positions of the spacecraft in Carrington coordinates
using the routine $get\_stereo\_lonlat$ provided by SolarSoftWare (SSW). The 3D density grid 
is then transformed from the Carrington heliographic system to the projected coordinate system 
viewed from STEREO-A or -B \cite{tho06}. Therefore, the LOS integral of $N(r)$ in Equation (\ref{eqpb}) 
becomes a simple summation in the $z$-direction (defined along LOS). 
Figure~\ref{fig:necr} shows the tomographic reconstructed 3D coronal electron density 
at 2 $R_\odot$. Figure~\ref{fig:nemap} shows the density distributions of the corona in the POS 
for COR1-A and -B at 12:00 UT on 8 February 2008. Figure~\ref{fig:pbmap} shows the corresponding
synthetic $pB$ images, which represent the ideal measurements without contamination by 
the scattered light and instrumental noises. 

\begin{figure}
  \centerline{\includegraphics[width=1.\textwidth,clip=]{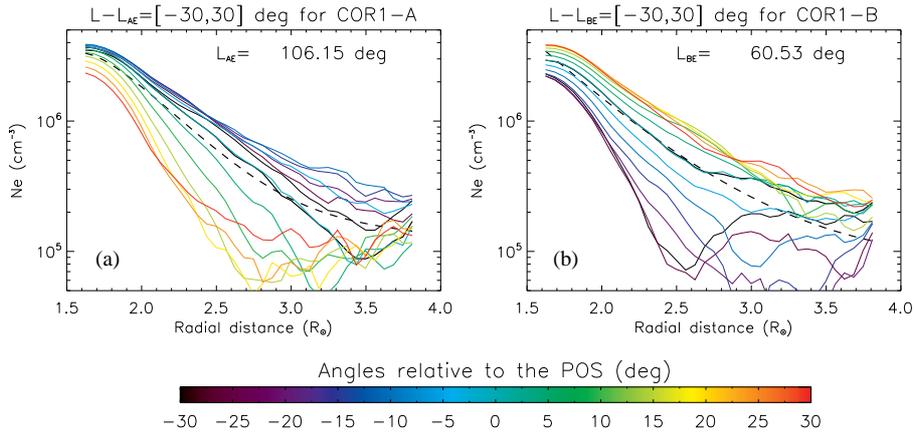}}
  \caption{Comparison of the SSPA density profile (dashed line) with the model
densities in 13 angular sections (solid lines). The thick black line 
stands for the case in the POS, while thin color lines for cases 
with angles relative to the POS to be $\pm$5, 10, 15, 20, 25,
30 degrees, where the negative value represents an angle measured
inward and the positive value an angle measured outward. (a) For COR1-A. (b) for COR1-B. 
$L_{\rm AE}$ and $L_{\rm BE}$ marked in (a) and (b) represent the longitude of the east limb 
in COR1-A and -B, respectively.  }
\label{fig:pbinm}
\end{figure}

\subsection{Radial Distribution of Electron Densities in Solar Minimum Streamer}
\label{sctstrm}
Coronal streamers are the most conspicuous, large-scale structures in the extended corona. 
The streamer belt is a usually continuous sheet of enhanced density associated with the magnetic
 neutral sheet or the current sheet (\opencite{sch73}; \opencite{guh96}; \opencite{sae07}; 
\opencite{kra09}, \citeyear{kra11}). Its shape gets 
progressively deformed from a rather flat plane at minimum solar activity to 
a highly warped surface at maximum solar activity ({\it e.g.}, \opencite{hu08}).
Some previous studies have suggested that the SSI assumptions are suitable to
symmetric streamers at low latitude, in particular, the solar minimum streamer belt
({\it e.g.}, \opencite{guh95}; \opencite{gib95}; \opencite{gib99}). 
Here we first test the SSPA inversion of the solar minimum streamer belt. A streamer belt 
during CR 2066 corresponding to solar minimum of Solar Cycles 23$/$24 is shown 
in Figure~\ref{fig:necr}. 

\begin{figure}
\centerline{\includegraphics[width=1.\textwidth,clip=]{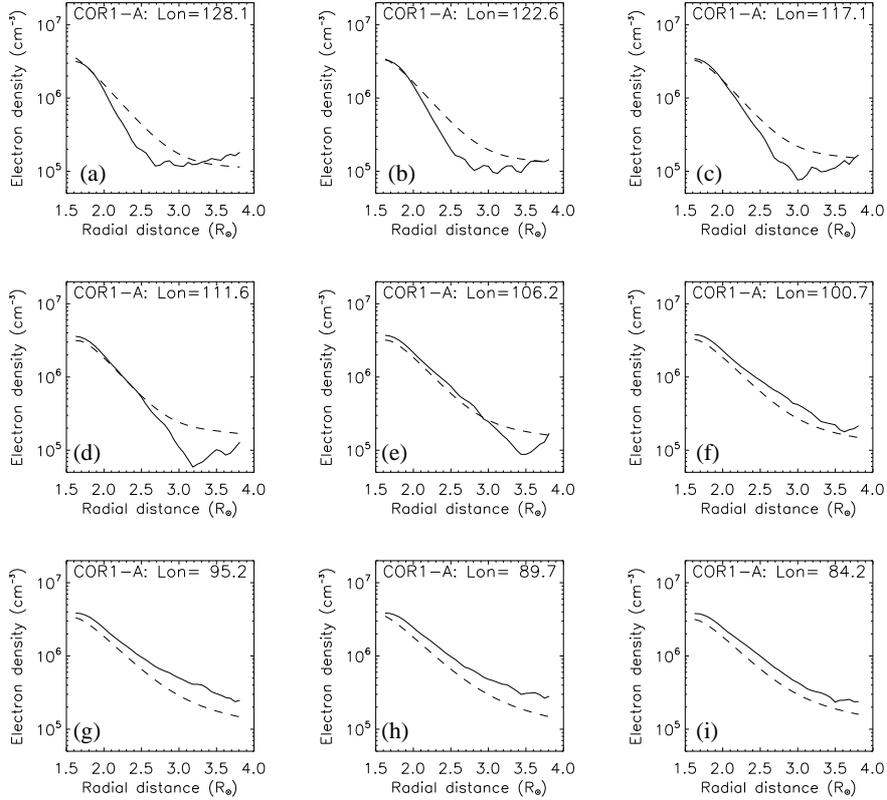}}
\caption{The electron density profiles derived by the SSPA inversion (dashed line)
as a function of longitudes. The analyzed radial trace is located at the position with 
Carrington latitude of 9$^{\circ}$ and longitude marked at the top of each panel and
shown with pluses in Figure~\ref{fig:necr}. The solid line in each panel 
represents the corresponding model densities in the POS. }
\label{fig:neva}
\end{figure}

For instance, we use the synthetic $pB$ images at 12:00 UT on 8 February 2008 when the separation 
angle between STEREO-A and -B was about 45$^{\circ}$. Figure~\ref{fig:pbmap} shows that the streamer 
belt on the east limb is almost edge-on, as inferred from the similar appearance in COR1-A and -B. 
In the edge-on condition, the coronagraph is looking along the streamer belt; {\it i.e.}, 
where all the streamers are at the same latitude behind each other along the LOS. 
In contrast, the streamer belt on west limb in COR1-B is face-on, showing a distinctly 
different shape from that in COR1-A. We choose a radial trace near the middle of the east-limb 
streamer (see marked positions in Figure~\ref{fig:pbmap}), where the $pB$ profiles 
for COR1-A and -B are nearly identical (Figure~\ref{fig:pblc}), and suppose that this location 
best meets the SSI condition. We then derive the radial distribution of electron density by
fitting the $pB$ data along this radial trace between 1.6 and 3.9 $R_\odot$ 
using both the SSPA method and the Van de Hulst technique ($pb\_inverter$ with $k$=5 in SSW) 
for an illustrated comparison, and show the inversion results in Figure~\ref{fig:pbinv}.
Note that since COR1 does not perform well at $\rho\lesssim$1.55 $R_\odot$
at some position angles due to interference from the occulter (see \opencite{fra12}), 
this led to defective $pB$ signals and thus unreliable reconstructions 
within this region, so we constrain all SSPA inversions in the following to regions with 
$\rho\ge1.6 R_\odot$. 

\begin{figure}
\centerline{\includegraphics[width=1.\textwidth,clip=]{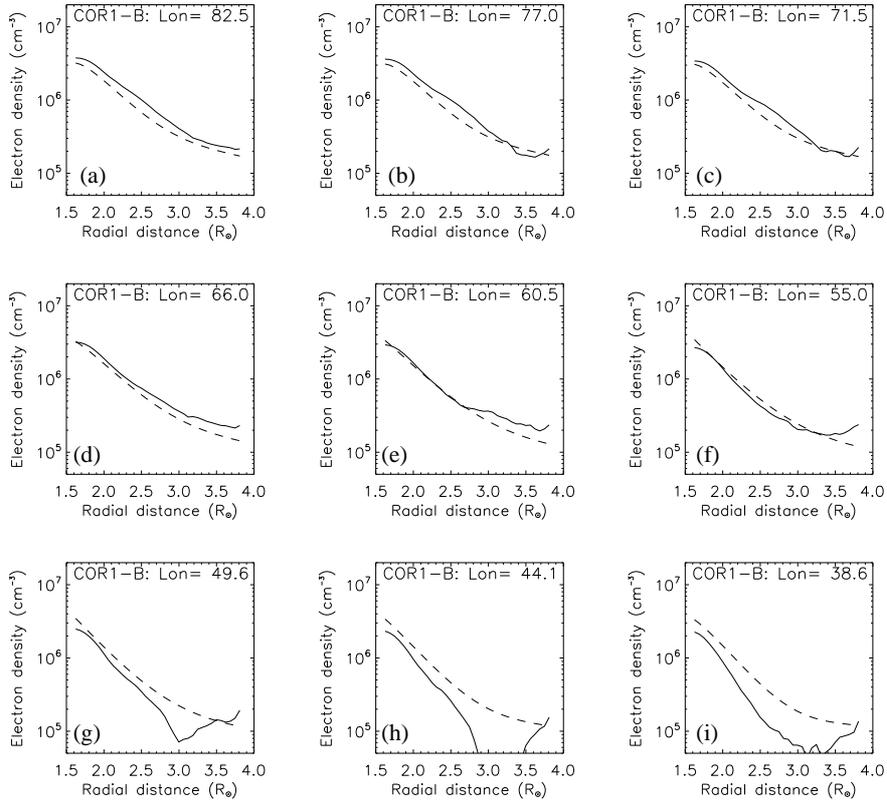}}
\caption{Same as Figure~\ref{fig:neva} but in the case for COR1-B. Locations 
of the analyzed radial trace are marked with diamonds in 
Figure~\ref{fig:necr}, whose latitude is 9$^{\circ}$ and longitude is marked at the top 
of each panel. The solid line represents the model densities in the POS
while the dashed line represents the SSPA densities.}
\label{fig:nevb}
\end{figure}

We find that the electron densities obtained from both $pB$ inversion methods agree well in 
radial distribution with the model 3D densities in the POS for selected longitudes 
corresponding to the edge-on streamer positions (see Figure~\ref{fig:pbinv} (c) and (d)). 
This confirms our theoretical prediction that the PPSA and the Van de Hulst inversions are equivalent. 
The average ratio of the SSPA to the model densities along the selected radial cut 
between 1.6 and 3.0 $R_{\odot}$ is 0.86$\pm$0.06 and 0.95$\pm$0.10 for COR1-A and -B, respectively.
Note that weak increase in density with height between 3.5 and 4.0 $R_{\odot}$
may be due to FOV effects (see discussion in Section~\ref{sctdc}). In addition, the lower panels 
of Figure~\ref{fig:pbinv} show that both the tomographic and SSPA density profiles determined
for streamers are comparable (with differences by $\sim$20\%) to the previous obtained by SSI method 
in the solar minimum \cite{gib99,sai77}. 

To examine to what extent the SSPA solution is consistent with the 3D density model,  
we compare the SSPA density profile with the model density profiles in 13 angular
sections to the POS within 30$^{\circ}$ in Figures~\ref{fig:pbinm}(a) and (b), which
show that the SSPA solution is closest to the distribution of 3D densities in the POS. 
This is as expected because the $pB$ integrals along the LOS are most heavily weighted toward
the regions near the POS (see \opencite{que02}; \opencite{fra10}).

\begin{figure}
\centerline{\includegraphics[width=1.\textwidth,clip=]{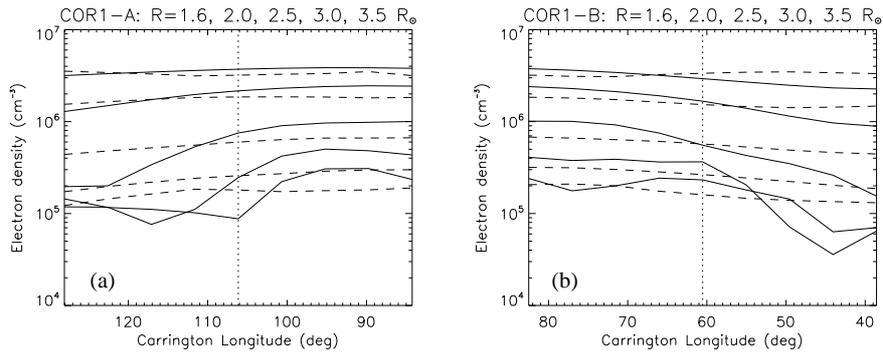}}
\caption{The coronal densities at 1.6, 2.0, 2.5, 3.0, and 3.5 $R_{\odot}$ as a function of 
Carrington longitudes, derived with the SSPA inversion (dashed lines) 
in comparison with those by the model in the POS (solid lines). 
The Carrington latitude of the analyzed radial trace is 9$^{\circ}$.
(a) For COR1-A. (b) For COR1-B. The vertical dotted lines indicate 
the middle position of the analyzed longitude range.}
\label{fig:netim}
\end{figure}

\begin{figure}
\centerline{\includegraphics[width=1.\textwidth,clip=]{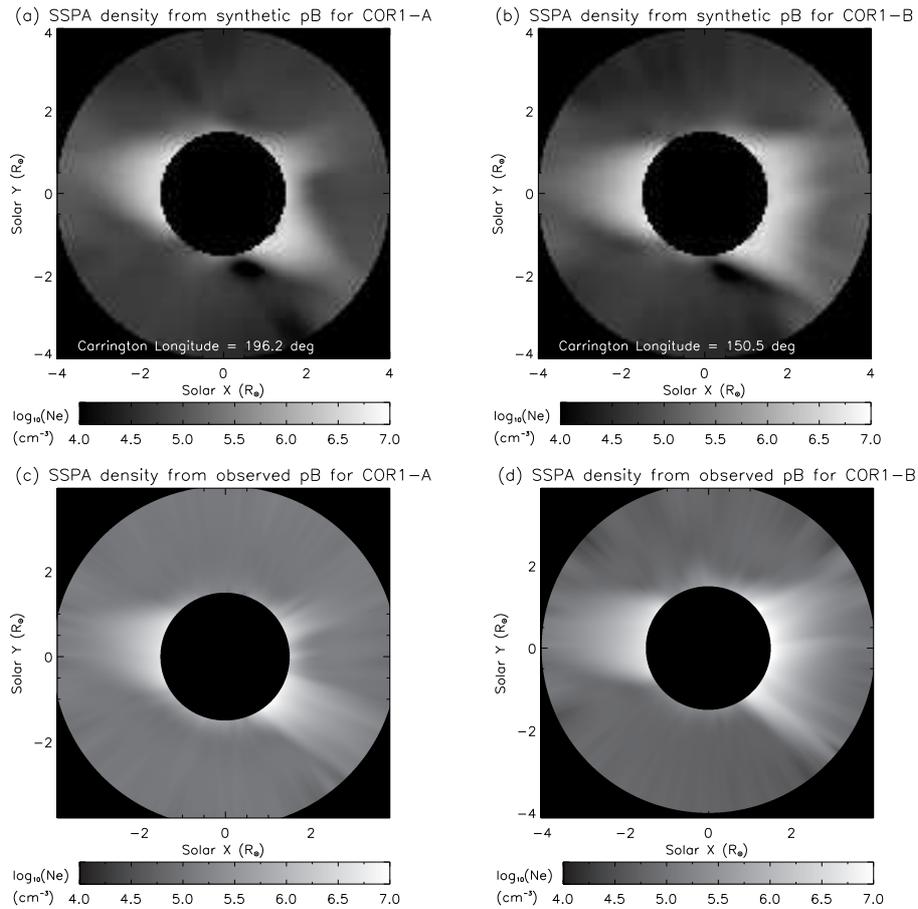}}
\caption{The 2D coronal density derived using the SSPA method from the synthetic $pB$ images,
corresponding to positions of COR1-A (a) and COR1-B (b) at 12:00 UT on 8 February 2008. 
Carrington longitudes of the viewing direction are marked at the bottom of the density maps. 
(c) and (d): Same as (a) and (b) but derived from the real observed $pB$ images 
at the same time. }
\label{fig:ne2d}
\end{figure}

\subsection{Longitudinal and Latitudinal Dependence of the Streamer Density}
\label{sctsdd}  
  We evaluate the SSPA inversion of the streamer as a function of longitude in the following. 
This corresponds to make the SSPA inversion of synthetic $pB$ images at different 
``virtual observing times". We synthesize the $pB$ images at nine times 
for COR1-A and -B, in a period from 20:00 UT, 6 February to 04:00 UT, 10 February with 
intervals of 10 hours, centered at 12:00 UT on 8 February 2008 (the time for the instance 
analyzed in the last section). The Sun rotated by 44$^{\circ}$ over this period, approximately 
equal to the separation angle between STEREO-A and B. So the ending-time location analyzed 
in COR1-A is approximately superposed with the starting-time location in COR1-B as shown 
in Figure~\ref{fig:necr}. We fit the $pB$ data along the same radial trace between 1.6 and
3.9 $R_{\odot}$ at these nine times using the SSPA method. Figures~\ref{fig:neva} 
and~\ref{fig:nevb} show comparisons of the obtained density profiles with the model densities 
in the POS for COR1-A and COR1-B, respectively. For the analyzed locations with longitude between 
$60^{\circ}$ and $106^{\circ}$ (approximately between the limbs $BE$ and $AE$ as shown 
in Figure~\ref{fig:necr}), we find that the density profiles derived by SSPA method have 
the similar shape as the model profiles over almost the whole FOV range (1.6--3.8 $R_{\odot}$), 
but the magnitudes are smaller than the model by about 20\%--50\%. The reason for good agreements 
during this period could be that the streamers in the LOS at the analyzed location are 
near the POS. The model density increases near the edge of FOV seen in 
Figures~\ref{fig:neva}(c)-(e) and Figures~\ref{fig:nevb}(g)-(i) likely result from 
the FOV effects in the tomographic inversion \cite{fra10}.
Figure~\ref{fig:netim} shows the SSPA density as a function of longitude at five heights 
(1.6, 2.0, 2.5, 3.0 and 3.5 $R_{\odot}$). The comparison with the model confirms 
the result above that the better agreement lies at the locations with the longitude 
in 60$^{\circ}$--106$^{\circ}$. We also find that the agreement tends to be better at lower heights. 
The average difference between the SSPA and model densities are $\sim$20\% for the region between 
1.6 and 2.0 $R_{\odot}$, while $\sim$40\%--50\% for 2.5--3.5 $R_{\odot}$.
The reason could be that the SSI condition is better meet at lower heights where the streamer
has a larger extent in the LOS, and thus the LOS integration in Equation (\ref{eqgktht}) over a
very long distance (compared to the width of the streamer) becomes more reasonable.   

Now we evaluate the latitudinal dependence of SSPA inversions of the corona. By assuming that 
the SSI condition holds locally for all angular positions around the Sun, the 2D coronal density
can be derived from a $pB$ image by fitting the radial profile at each angular position 
using the SSPA method. For the case analyzed in Section~\ref{sctstrm}, we reconstruct 2D 
coronal densities in the POS by fitting radial profiles between 1.6 and 3.7 $R_{\odot}$ for 360
position angles with intervals of 1$^{\circ}$ from both synthetic and observed $pB$ images,
and show the results in Figure~\ref{fig:ne2d}. Compared to the tomographic densities in the POS 
(see Figure~\ref{fig:nemap}), the SSPA coronal densities determined
from the observed images do not show very low-density regions around streamers, while those 
from the synthetic images have a small region of zero (or negative) densities 
at the southern side of the west-limb streamer, where the synthetic $pB$ signals are very low 
(below the background level in non-streamer regions). The zero density regions
in tomographic reconstructions could be caused either (or both) due to coronal dynamics 
or due to real very small density in these coronal regions (see discussions 
in Section~\ref{sctdc}). 

\begin{figure}
\centerline{\includegraphics[width=1.\textwidth,clip=]{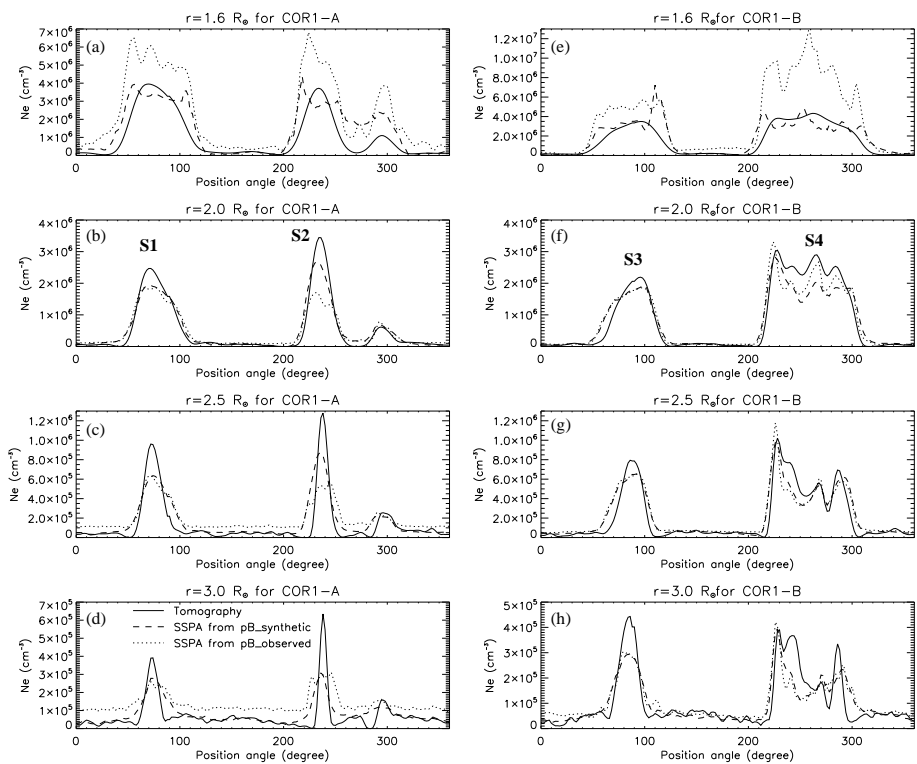}}
\caption{The coronal densities along four circular paths at 1.6 (a), 2.0 (b), 2.5 (c), 
and 3.0 $R_{\odot}$ (d) for COR1-A. The position angle is counted anticlockwise relative 
to the north pole. For each circle its eastern part (position angles 0$^{\circ}$--180$^{\circ}$)
has the longitude of 106.2$^{\circ}$, and its western part has the longitude of 286.2$^{\circ}$.
The solid line represents the model densities, and the dashed line
and the dotted line represent the SSPA densities from the synthetic
and observed $pB$ images, respectively. (e)-(h): Same as panels (a)-(d) but for COR1-B.
For each circle its eastern part has the longitude of 60.5$^{\circ}$, and its western part 
has the longitude of 240.5$^{\circ}$.}
\label{fig:neap}
\end{figure}

For a quantitative comparison, we plot the SSPA and model density profiles as a function 
of position angles at four heights (1.6, 2.0, 2.5, and 3.0 $R_{\odot}$) in Figure~\ref{fig:neap}.
For the three edge-on streamers marked with S1--S3, we measure their peak densities 
and angular widths (in FWHM). The ratio of peak densities from the SSPA to the model is 
on average 0.82$\pm$0.06, 0.72$\pm$0.09, and 0.62$\pm$0.12, and the ratio of angular widths 
is on average 1.24$\pm$0.02, 1.67$\pm$0.12, and 1.90$\pm$0.27 at 2.0, 2.5, and 3.0 $R_{\odot}$, 
respectively. The results indicate that the SSPA inversion underestimates the peak density 
of streamers by about 20\%--40\%, while overestimates the angular width by about 20\%--90\%. 
The increase of deviations with height in the peak density of streamers 
by the SSPA method from the model may be caused by narrowing of the streamer (belt) 
with height (see Figure 2 of \opencite{kra09}).
As the streamer width along the LOS decreases with the height, the condition for
the SSI assumption becomes worse. The reason for the latitudinal spreading of the SSPA density
relative to the model may lie in that the analyzed streamers are not exactly edge-on, 
{\it i.e.}, the streamer belt at the analyzed locations is actually tilted somewhat 
away from the equatorial direction (or the LOS near the limb, see Figure~\ref{fig:necr}). 
In such a case at the locations near the
edge of streamer in the $pB$ image the contribution is larger from points along the LOS
behind or in front of the POS, leading to an overestimation of the model density in the POS
by the SSPA inversion at these places. In addition, we notice for the face-on streamer 
(marked S4) in COR1-B, its SSPA density profiles are consistent with the model 
densities as well, this is because this streamer is by coincidence located near 
the POS at this instance (Figure~\ref{fig:necr}).
 
In Figure~\ref{fig:neap}, we also compare the SSPA densities inverted from the synthetic 
$pB$ image with those from the observed $pB$ image, and find that 
they are generally consistent except in the region near the occulter of COR1, where the
SSPA inversions from the observed $pB$ are much larger ($\sim$2--3 times) than the model 
densities (see panels (a) and (e) in Figure~\ref{fig:neap}). The version of tomographic 
reconstruction used here most likely underestimated the density there as the solution at 
grid points close to the occulter is less constrained by the observational data.  

\begin{figure}
\centerline{\includegraphics[width=1.\textwidth,clip=]{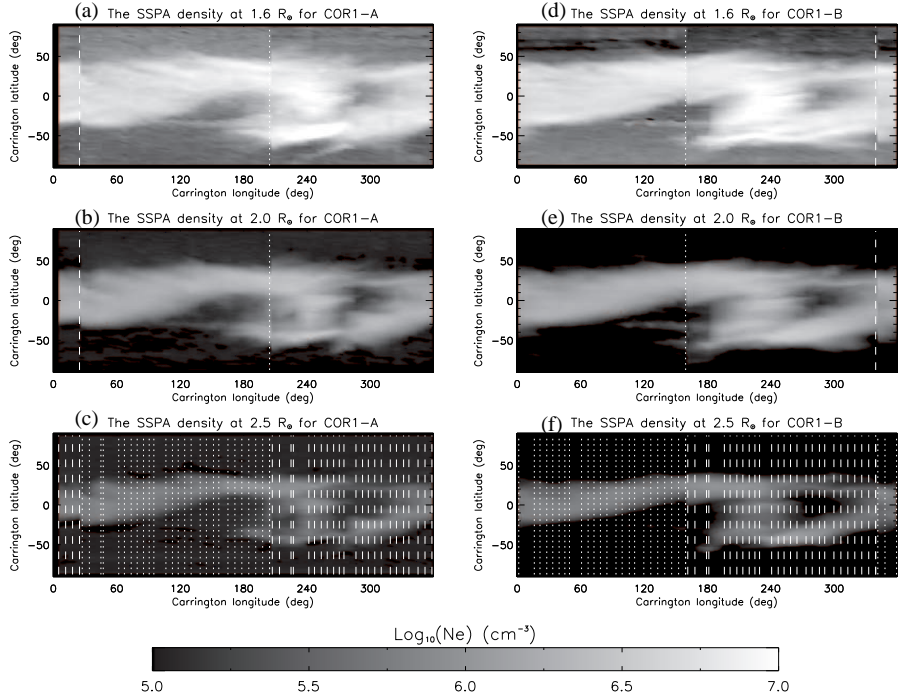}}
\caption{The SSPA-reconstructed 3D coronal densities from the two-week dataset of COR1 $pB$ images. 
(a)-(c): The spherical cross sections at 1.6, 2.0, and  2.5 $R_{\odot}$ for COR1-A, and  
(d)-(f) for COR1-B. In panels (a), (b), (d) and (e), the dotted line and the dashed line 
represent the positions of the east limb and the west limb, respectively, 
of the earliest-observed $pB$ image (at 00:25 UT, 1 February 2008).
In panels (c) and (f), the vertical dotted and dashed lines represent the positions of 
the east limb and west limb, respectively, for all $pB$ images used in the reconstructions. }
\label{fig:synab}
\end{figure}

\begin{figure}
\centerline{\includegraphics[width=1.\textwidth,clip=]{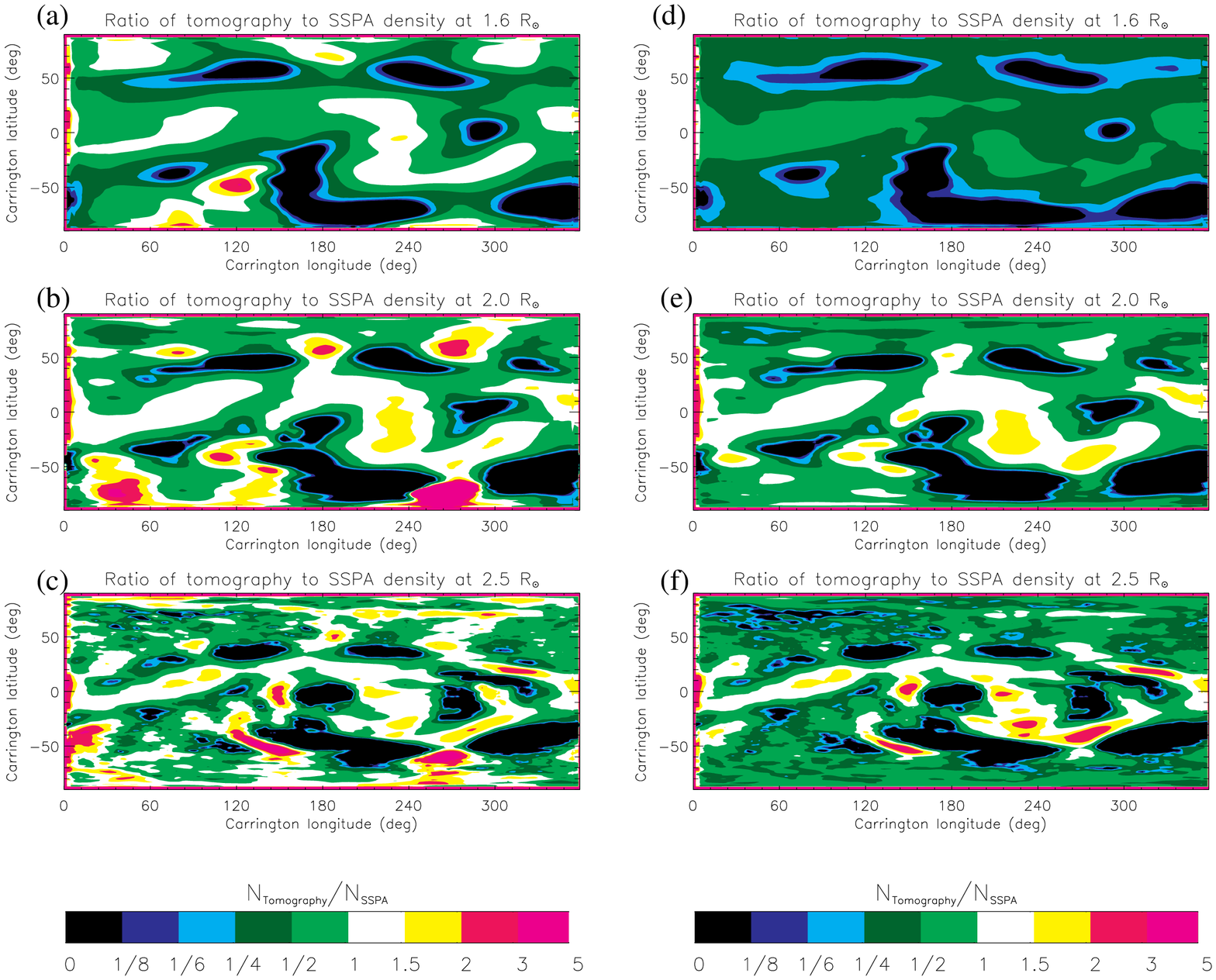}}
\caption{The ratio of tomographic density to the SSPA average density for COR1-A and -B 
at 1.6, 2.0, and 2.5 $R_{\odot}$ for two cases. (a)-(c): The case for the SSPA 3D coronal
density obtained using synthetic $pB$ data based on the 3D density model. 
(d)-(f): The case for the SSPA 3D coronal density obtained using the real $pB$ data observed 
by COR1-A and -B.  The colors represent the ratio values in a range 0--5 
(see color bars at the bottom). A version with continuous color scale is available as 
Electronic Supplementary material.}
\label{fig:syncp}
\end{figure}

\section{Reconstruction of the 3D Coronal Density using SSPA Method for CR 2066}
\label{sctd3d}
In the above sections, we evaluated the SSPA method by comparing the derived density distribution
as a function of radial distance, longitude and latitude with the 3D tomographic model.
The assessments indicate that the SSPA inversion can determine the 2D coronal density from
a $pB$ image, which approximately agrees with the 3D tomographic densities for streamers near the POS. 
This suggests that we may reconstruct a 3D density of the corona by applying the SSPA inversion 
to a 14 day data set from COR1-A or -B. In this section, we demonstrate the SSPA 3D density
reconstructions using the same data set (consisting of 28 $pB$ images from COR1-B) as used for 
the reconstruction of 3D tomographic model, and the simultaneous COR1-A data set. 
The data cadence of about 2 images per day corresponds to a longitudinal step of about $6^\circ$.
First we determine the 2D density distribution by fitting the radial $pB$ data between 
1.6 and 3.7 $R_{\odot}$ using the SSPA inversion at 142 angular positions 
(with intervals of 2.5$^{\circ}$) surrounding the Sun for each image. Then we map the 
east-limb and west-limb density profiles of all images to make a synoptic map at a certain height,
based on their Carrington coordinates  (neglecting the inclination of the solar rotational axis to 
the ecliptic). A 3D density reconstruction is made of 25 synoptic maps for the radial heights 
from 1.5 to 3.9 $R_{\odot}$ with an interval of 0.1 $R_{\odot}$. For each synoptic map,
we convert the irregular grid into the regular grid using the IDL function, {\it trigrid}. 
Although the 3D density reconstruction can be made using the SSPA method from the COR1 data 
with a cadence as high as 5 minutes, the higher temporal resolution may not help improve 
its actual angular resolution in longitude due to intrinsic limitations of the SSPA method 
(see Appendix).

Figure~\ref{fig:synab} shows the SSPA-reconstructed 3D coronal density at 1.6, 2.0 and 2.5
$R_{\odot}$ for COR1-A and -B. Some discontinuities of the density are seen at two longitudes 
which separate the two regions made of the east-limb and west-limb inversions. 
These flawed structures may be due to temportal changes of streamers and/or the effect 
due to neglecting the solar axial tilt in the reconstruction. We smooth these discontinuities 
by averaging the two reconstructions from COR1-A and COR1-B and then make 
a $10^{\circ}\times10^{\circ}$ smoothing. In Figure~\ref{fig:syncp}, we show the ratio of 
tomographic density to the SSPA average density for the COR1-A and -B reconstructions with
smoothing for two cases, one for the SSPA densities obtained using the synthetic data 
(panels (a)-(c)), and the other for the SSPA densities obtained using the real data 
(panels (d)-(f)).  Both cases indicate that the density ratios
in the streamer belt are very close to 1, within a factor of two or so ($\sim$0.5--1 
at 1.6 $R_\odot$, $\sim$1--2 at 2.0 $R_\odot$, and $\sim$1--3 at 2.5 $R_\odot$).
The good consistency validates that the SSI assumptions are very appropriate for the 
streamer belt in the solar minimum. 

\begin{figure}
\centerline{\includegraphics[width=0.8\textwidth,clip=]{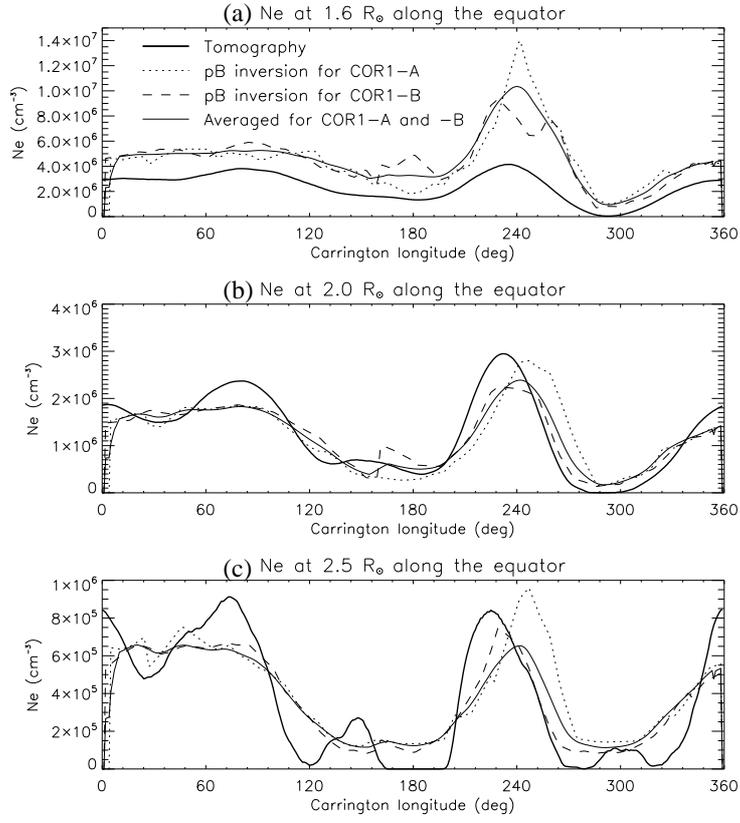}}
\caption{Comparison of the SSPA density profiles with the tomographic density profiles 
(thick solid line) along the equator at (a) 1.6, (b) 2.0, and (c) 2.5 $R_{\odot}$. 
The dotted and the dashed lines represent the SSPA coronal densities for COR1-A and -B, 
respectively, while the thin solid line represents their average with a 10-degree smoothing.}
\label{fig:synpf}
\end{figure}

\begin{figure}
\centerline{\includegraphics[width=1.\textwidth,clip=]{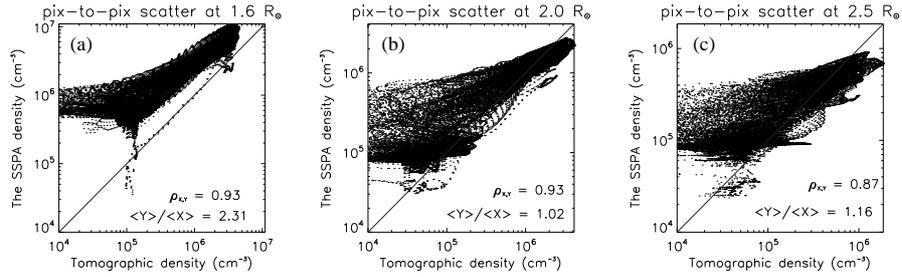}}
\caption{The pixel-to-pixel scatter plots of the SSPA 3D coronal density (averaged for COR1-A and
-B with a 10-degree smoothing) versus the tomographic 3D coronal density, 
at (a) 1.6, (b) 2.0, and (c) 2.5 $R_{\odot}$. The linear Pearson correlation coefficient 
($\rho_{X,Y}$) and the ratio of the average densities ($<Y>/<X>$) are marked in each plot. }
\label{fig:pxsct}
\end{figure}

For a quantitative comparison between the tomography and the SSPA densities (obtained from real 
data), we also show their density profiles along an equatorial cut in Figure~\ref{fig:synpf}. 
We find that they best match at 2 $R_\odot$. The same 
result is also indicated by the pixel-to-pixel scatter plots in Figure~\ref{fig:pxsct}.
We obtain the ratio of the SSPA to the tomographic average density is 1.02, and their 
linear Pearson correlation coefficient is 0.93 for the reconstructions at 2 $R_\odot$. 
In addition, the scatter plots show that the dispersion is larger for those pixels
with smaller values, where the SSPA densities are much larger than those obtained 
by the tomography.

\section{Discussion and Conclusions}
\label{sctdc}
In this study, we have, for the first time, evaluated the SSI method using the 3D model 
of the coronal electron densities reconstructed by tomography from the STEREO/COR1 $pB$ data.
Our study is instructive for more efficient use of the SSI technique to invert the $pB$
observations from ground- and space-based coronagraphs, in particular, the COR1 data.
We demonstrate both theoretically and observationally that the SSPA method and 
the Van de Hulst inversion are equivalent SSI techniques when the radial densities or 
$pB$ signals are assumed in the polynomial form of high degrees (more than two). 
The polynomial degree of five is suitable for COR1 data inversions. Thus, assessment 
results of the SSPA method can also be applied to the Van de Hulst inversion technique. 
We determine radial profiles of the streamer density from the COR1-A and -B synthetic 
$pB$ images as well as their longitudinal and latitudinal dependencies. 
We find that the SSPA density values are close to the model for the core of streamers 
near the POS, with differences (or uncertainties if we regard the model input 
as a true solution) typically within a factor of $\sim$2.
This result is consistent with those evaluated using UV spectroscopy \cite{gib99, lee08}. 
We find that the SSPA density profiles tend to better match the model at lower heights 
($\lesssim$ 2.5 $R_\odot$). Our results confirm the suggestion in some previous studies 
that the SSI assumption is appropriate for the edge-on streamers or the streamer belt 
during the solar minimum. We suggest that the edge-on condition for streamers may be 
determined by tomography method or by examining the
consistency between simultaneous $pB$ measurements from COR1-A and -B in radial distribution, 
when the two spacecraft have a small angular separation ({\it e.g.}, less than 45$^{\circ}$).
We also find the SSPA streamer densities are more spread in both longitudinal and 
latitudinal directions than in the model.

We demonstrate the application of the SSPA inversion for reconstructions of the 3D coronal 
density near the solar minimum, and show that the SSPA 3D density for the streamer belt is 
roughly consistent in both position and magnitude with the tomographic reconstruction. 
The synoptic density maps derived by the SSPA method show some discontinuities 
at the longitudes that separate the regions made of the east- and west-limb inversions. 
These discontinuities may be due to temporal changes of coronal structures and/or 
the effect of tilt of the solar rotation axis on the poles' visibility that is not considered. 
They can be smoothed out during post-processing by smoothing and averaging 
the density distributions from COR1-A and -B, but such treatments 
will reduce the spatial resolution. In comparison, the tomographic inversion
can fully take into account the tilt effect of the solar pole and produce the density 
distributions smoother in these discontinuity regions \cite{fra02, kra09}. We estimate that
the SSPA method may resolve the coronal density structure near the POS with an angular
resolution of $\sim$50$^{\circ}$ in longitudinal direction. Given this limitation, 
the SSPA reconstruction using $pB$ data with higher cadence ({\it e.g.}, more than 
three images per day) would  not help improve its actual angular resolution.

Although the current state of the tomographic method has allowed to routinely obtain 
the 3D coronal densities, we speculate that the SSI method could be complementary to 
the tomography when used for the interpretation of observations in such cases as during 
maximum of solar activity, in some regions where tomography gives zero density values, 
or in the regions near the edges of FOV. 

The zero density values in the tomographic reconstruction (so called ``zero-density 
artifacts", ZDAs) could be caused either (or both) due to coronal dynamics 
\cite{fra02, fra07, vas08, but10}, or due to real very small density in these coronal regions 
which are below the error limit in the tomography method \cite{kra09}. The latter reason 
is also supported by results of MHD modeling \cite{air11}. In the former case, 
the SSI method could be complementary to tomography, while in the latter case the SSPA method 
gives much larger values than in the tomographic model. The FOV artifacts in the tomography 
are due to the finite coronagraph FOV that causes the reconstructed density to increase at 
the regions close to the outer reconstruction domain \cite{fra10, kra09}. 
However, this FOV artifacts can be reduced by extending the outer reconstruction domain 
beyond the coronagraph FOV limit \cite{fra10, kra13}. Another way to obtain more correct 
density values in this region could be by using the SSPA method which does not imply 
strict outer boundaries for LOS integration. Thus the estimate of uncertainties of 
the SSPA method should be limited to the regions with distances less than about 3.5 $R_{\odot}$ 
for the tomographic model used. Also, the use of MHD model in order to produce artificial 
data can be useful for this test. However this will be a subject of future research.

The tomography generally assumes that the structure of the corona is stable over 
the observational interval, {\it e.g.}, two weeks of observations made by a single spacecraft, 
although for some coronal regions that are exposed to the spacecraft for only about a week 
during the observation the stationary assumption can be reduced to about a week \cite{kra11}. 
However, such an assumption is hard to meet during solar maximum
or times of enhanced coronal activity. The CME catalog in the NASA CDAW data center 
shows that the CME occurrence rate increases from $\sim$0.5 per day near solar minimum 
to $\sim$6 near solar maximum during Solar Cycle 23 \cite{gop03, yas04}. 
Although our assessment results for the SSPA method are based on a static coronal model, 
their validity may not be limited to the static assumption. Because the key factor for 
the SSPA inversion for obtaining a good estimate of the 2D coronal density is 
an instantaneous (local) symmetric condition for coronal densities along the LOS. 
The minimal size of this local symmetry is limited by the angular resolution in longitude 
which is about $50^\circ$. Therefore, the SSI method can be used to estimate the density of
a dynamic coronal structure in terms of weighted average over the region with this angular
size in longitude. For this reason it may be a better choice to use a combination of the 
tomography and the SSI inversions for interpretations of radio bursts and shocks produced 
by CMEs in the case when coronal structures of interest evolve 
quickly with time \cite{lee08, she13, ram13}. In addition, the SSI method is also
often applied to the cases when observational data are not suitable for tomography,
{\it e.g.}, solar eclipses.

The 3D MHD models of the corona using the synoptic photospheric magnetic field data 
have been successfully used to interpret solar observations, 
including total eclipses and ground-based ({\it e.g.} MLSO/MK4) 
images of the corona ({\it e.g.}, \opencite{lin99}; \opencite{mik99}). We suggest that 
the evaluation of the SSI method based on such a global MHD model may be necessary in the future.
The profits using such MHD models to estimate the uncertainties of the SSI method
could be in avoidance of the ZDA and FOV effects. The modeled corona also allows evaluations of 
the SSI method down to very low heights ({\it e.g.}, the region between 1.1 and 1.5 $R_{\odot}$) 
where the corona is much more structured. Moreover, a simulated time dependent corona 
from a time-evolving MHD model would allow us to estimate the uncertainties in tomography 
and the SSI inversion when they are applied to the dynamic corona, especially during 
the solar maximum. This needs detailed investigations in the future.

\appendix

{\bf Estimates of Angular Resolution of the SSPA Method in Longitude}\\

\begin{figure}
\centerline{\includegraphics[width=1.0\textwidth,clip=]{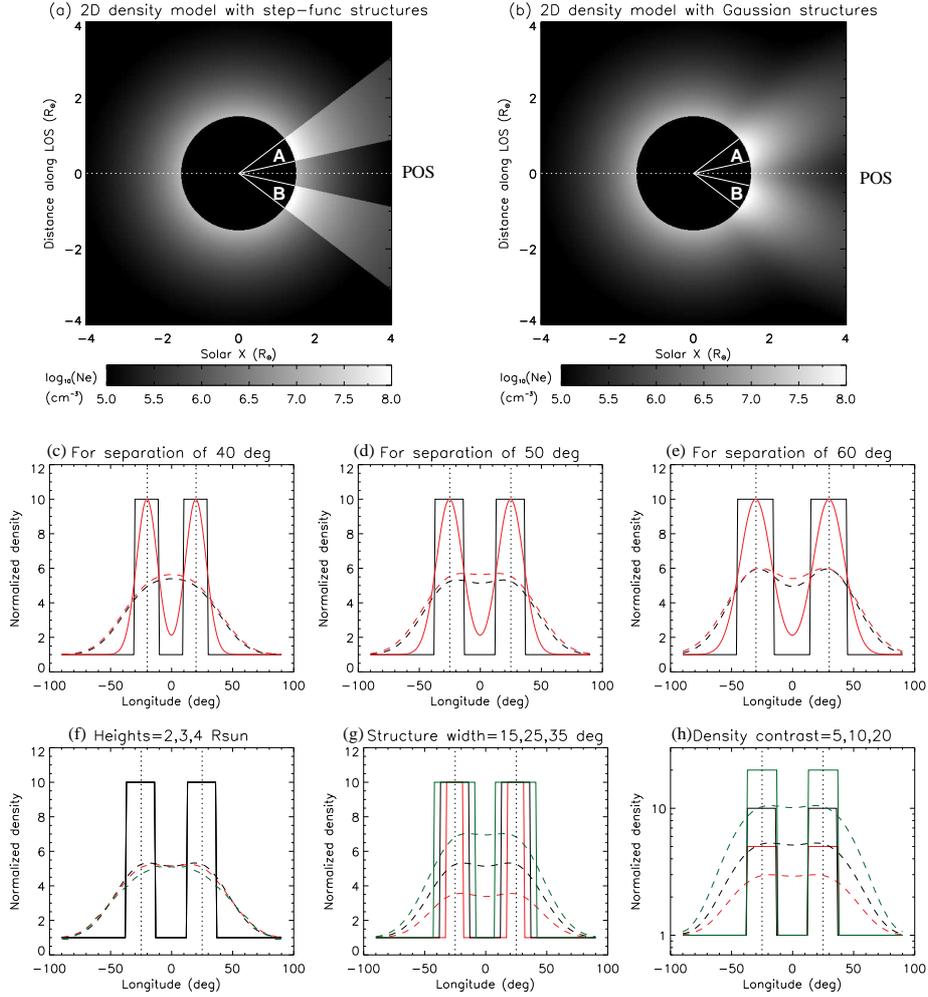}}
\caption{2D density models of the corona from 1.5 to 4.0 $R_{\odot}$ in the equatorial 
plane with two structures (marked A and B) of (a) the step profile and (b) the Gaussian 
profile in longitude. The horizontal dotted line represents the POS. (c)-(e): Comparisons of
the SSPA density (in dashed line) with the model density (in solid line) in the POS 
at 2 $R_{\odot}$ as a function of longitudes for the two structures with angular separation 
of 40$^{\circ}$ (c), 50$^{\circ}$ (d), and 60$^{\circ}$ (e). The black curves represent 
the case for the model density with the step profile, and the red curves the case for
the model density with the Gaussian profile. 
(f): The comparison between the SSPA and model densities at different heights. 
The black, red, and green dashed lines represent the SSPA density profiles 
at 2, 3 and 4 $R_{\odot}$, respectively. Note that the model density profiles (in solid line),
normalized to the coronal background, are same for the three heights.
(g): The case for the structures with different widths of 15$^{\circ}$ (red curves),
25$^{\circ}$ (black curves), and 35$^{\circ}$ (green curves). The solid lines represent
the model density in the POS, and the dashed lines the SSPA density. (h): The case for the
structures with different density contrast ratios to the background of 
5 (red curves), 10 (black curves), and 20 (green curves). The solid lines represent
the model density in the POS, and the dashed lines the SSPA density. 
In (f)-(h), except for one parameter that is set as different values, the other parameters 
are set to be 50$^{\circ}$ for angular separation, 2 $R_{\odot}$ for heliocentric height, 
25 $^{\circ}$ for structure width, and 10 for density contrast ratio. 
The vertical dotted lines in (c)-(h) mark the central position in the structure.}
\label{fig:reso}
\end{figure}

To determine the angular resolution of the SSPA method in the longitudinal direction, 
a numerical experiment is performed using 2D coronal models.  We construct a 2D density
model by first using the \inlinecite{sai77} equatorial background density model to build 
a background corona of rotational symmetry in the equatorial plane, and then inserting 
two structures into it. The structures have the angular width of $\phi$, the density 
contrast ratio of $d$ to the background, and the longitudinal profile following 
a step function or Gaussian function. For the Gaussian-type structure, its FWHM is 
set as $\phi$. Figures~\ref{fig:reso}(a) and (b) show the two types of 2D coronal models 
with about the same FOV as COR1, where two structures are separated by an angle of 
2$\phi$, thus the width of the gap between them is also equal to $\phi$ in the step-profile
case. We assume that the two structures (marked A and B) are located at the longitudes of 
$\phi$ and $-\phi$, respectively, i.e. defining their middle position as the origin of 
longitude. So for the cases shown in Figures~\ref{fig:reso}(a) and (b), the origin of
longitude is just located in the POS at the west limb. 

We synthesize $pB$ data for the west-limb region from 1.5 to 4 $R_{\odot}$ at longitudes 
in the range from $-90^{\circ}$ to 90$^{\circ}$ using Equation~(\ref{eqpb}), and then derive 
the electron density using the SSPA method by fitting the synthetic $pB$ data. 
The panels (c)-(e) show comparisons of the SSPA density with the model 
density in the POS at 2 $R_{\odot}$ as a function of longitudes for different angular 
distances between two artificial structures with the density contrast ratio $d$=10. 
We find that for both types of the structure (in the step or Gaussian profile), the minimum 
resolvable distance (or angular resolution) is about 50$^{\circ}$. To examine the
effects of radial distance, structure width, and density contrast on the obtained resolution,
we make a parametric study, and show the results in panels (f)-(h). We find that the
resolution is only slightly dependent on the radial distance and structure width, which
is better for the lower heights and narrower widths, but almost independent on the density
contrast. In addition, we also find that the obtained SSPA peak density is about a half 
of the model density for the analyzed structures in most of the cases, which is consistent with our
results for edge-on streamers. These numerical experiments suggest that for coronal structures 
with more smoothing profile and larger extension in the longitudinal direction and 
with lower density contrast to the background, which can be regarded as better conditions 
meeting the spherically symmetric assumption, the SSPA solutions are more accurate.

%
\begin{acks}
The work of TW was supported by the NASA Cooperative Agreement
NNG11PL10A to the Catholic University of America and NASA grant NNX12AB34G. 
We very much appreciate to Dr. Maxim Kramar for his suggestions that led to an
improved estimation of angular resolution of the SSPA method in Appendix. 
We also thank the anonymous referee for his/her valuable comments in 
improving the manuscript.

\end{acks}


\end{article} 
\end{document}